# Accurate Correlation Energy Functional for Uniform Electron Gas from an Interpolation Ansatz without Fitting Parameters

Qing-Xing Xie,[1] Jiashun Wu[1] and Yan Zhao[1,2*]

[1]The Institute of Technological Sciences, Wuhan University, Wuhan 430072, People's Republic of China.

[2]State Key Laboratory of Silicate Materials for Architectures, International School of Materials Science and Engineering, Wuhan University of Technology, Hubei, Wuhan 430070, People's Republic of China

*Corresponding author, E-mail: yan2000@whut.edu.cn

**Abstract**

We report an analytical representation of the correlation energy $e_c(r_s, \zeta)$ for a uniform electron gas (UEG), where $r_s$ is the Seitz radius or density parameter and $\zeta$ is the relative spin polarization. The new functional, called W20, is constructed to capture the known high-density ($r_s \to 0$) and low-density ($r_s \to \infty$) limits (for $\zeta = 0$ and 1) without any fitting parameters. The comparative assessment against the recent quantum Monte Carlo (QMC) results shows that the performance of the W20 functional is comparable to the popular parametrized UEG correlation functionals. On average, W20 agrees with QMC and PW92 [Phys. Rev. B 45, 13244 (1992)] within 0.01 eV for $0.5 \leq r_s \leq 20$, and W20 recovers the correct high- and low-density limits, whereas the QMC-data fitted UEG correlation functionals do not recover the low-density limit. W20 shows better accuracy than the popular QMC-data fitted UEG correlation functionals for the low-density region.

# I. INTRODUCTION

Since the birth of density functional theory (DFT) [1,2], the hypothetical uniform electron gas (UEG) has served as a ubiquitous model for the developments of density functionals, and all modern DFT calculations involve UEG-based components in the underlying exchange-correlation functionals. Although the analytical representation of the UEG exchange energy is known [3], the UEG correlation energy per electron, $e_c(r_s, \zeta)$, does not have a closed analytical form. Here, the Seitz radius, $r_s$, is related to the electron density $\rho$ as

$$\rho = \rho_\uparrow + \rho_\downarrow = \left[\frac{4\pi}{3}(r_s a_B)^3\right]^{-1}, \tag{1}$$

where $\rho_\uparrow$ and $\rho_\downarrow$ are the up- and down-spin electron densities, respectively. $a_B$ is the Bohr radius $=\hbar/(me^2)$, and the atomic units are employed ($\hbar = m = e^2 = 1$). $\zeta$ is the relative spin polarization,

$$\zeta \equiv \frac{\rho_\uparrow - \rho_\downarrow}{\rho}. \tag{2}$$

The widely used VWN80 [4], PZ81 [5], and PW92 [6] UEG correlation functionals have been developed by parametrizing and interpolating between the discrete quantum Monte Carlo (QMC) data of Ceperley and Alder [7]. Inspired by the Wigner [8] interpolation, Sun, Perdew, and, Seidl [9] recently proposed a density parameter interpolation (DPI) between the high-density ($r_s \to 0$) and low-density ($r_s \to \infty$) limits of the UEG $e_c(r_s, \zeta)$ without fitting to the QMC data. Later, Loos and Gill [10] have found that one of the expansion coefficients used in the original DPI model was in error, and Bhattarai et al. [11] corrected this error and compared the two DPI functionals to VWN80, PZ81, PW92, and the QMC data of Spink et al. [12], and they found that the corrected DPI model agrees with the original DPI within 0.001 eV, showing the robustness of the DPI approach.

In 2016, Chachiyo[13] developed an elegant functional (it is named as C16 in this work) based on the second-order Moller-Plesset perturbation theory, and is of the form:

$$e_c(r_s, \zeta = 0, 1) = a(\zeta)\ln(1 + \frac{b(\zeta)}{r_s} + \frac{b(\zeta)}{r_s^2}) \quad (3)$$

The parameters $a(\zeta)$ and $b(\zeta)$ ($\zeta = 0, 1$) in Eq. (3) have been derived from the leading coefficients $a_0(\zeta)$ and $b_0(\zeta)$ ($\zeta = 0, 1$) in the high-density asymptotic expansion of $e_c(r_s, \zeta)$ [9,14-17]:

$$\begin{aligned} e_c(r_s, \zeta) &= \sum_{n=0}^{\infty}[a_n(\zeta)\ln(r_s) + b_n(\zeta)]r_s^n \\ &= a_0(\zeta)\ln(r_s) + b_0(\zeta) + r_s[a_1(\zeta)\ln(r_s) + b_1(\zeta)] + ... \quad (r_s \to 0) \end{aligned} \quad (4)$$

Later, Karasiev [18] introduced two empirical parameters into Eq. (3) by using the QMC data at $r_s = 50$ (this model is named as revC16 in this work), and he showed that the performance of revC16 is better than C16 for the predictions of UEG correlation energies.

Due to the limited flexibility of Eq. (3), the C16 and revC16 models can only recover the leading $a_0(\zeta)$ and $b_0(\zeta)$ ($\zeta = 0, 1$) coefficients in Eq. (4), but not the $a_1(\zeta)$ and $b_1(\zeta)$ ones. Furthermore, neither of the two models recovers the low-density limit expansion of $e_c(r_s, \zeta)$ [4,6,8,9,11,19]:

$$\begin{aligned} e_c(r_s, \zeta) &= \frac{f_0 - c_x(\zeta)}{r_s} + \frac{f_1}{r_s^{3/2}} + \frac{f_2 - c_s(\zeta)}{r_s^2} \\ &+ \sum_{n=3}^{\infty} \frac{f_n}{r_s^{1+n/2}} + e_{\exp}(r_s, \zeta) \quad (r_s \to \infty) \end{aligned} \quad (5)$$

where $f_0$, $f_1$, and $f_2$ are constant coefficients; $c_x(\zeta)$ and $c_s(\zeta)$ are spin scaling functions for the UEG exchange energy and noninteracting kinetic energy, respectively [6,9]:

$$c_s(\zeta) = \frac{3}{10}\left(\frac{9\pi}{4}\right)^{2/3}\frac{1}{2}\left[(1+\zeta)^{5/3} + (1-\zeta)^{5/3}\right], \quad (6)$$

$$c_x(\zeta) = -\frac{3}{4\pi}\left(\frac{9\pi}{4}\right)^{1/3} \frac{1}{2}\left[(1+\zeta)^{4/3} + (1-\zeta)^{4/3}\right]. \tag{7}$$

## II. INTERPOLATION ANSATZ

The aim of this work is to construct an accurate UEG correlation functional by interpolating between the high-density ($r_s \to 0$) and low-density ($r_s \to \infty$) limits of the UEG $e_c(r_s, \zeta)$ with a new interpolation ansatz without fitting to the QMC data. The proposed functional will be referred to as "W20" (stands for Wuhan 2020). The W20 model recovers both the high- and low-density limits (Eqs. (4) and (5) for $\zeta = 0, 1$). The exact or near-exact coefficients in Eqs. (4) and (5) are listed in TABLE I and are employed in W20. The derivations of these coefficients have been summarized in the advanced review of UEG by Loos and Gill [20].

In order to achieve the high- and low-density asymptotic behaviors of Eqs. (4) and (5), we choose

$$e_c(r,\zeta) = -\frac{a_0(\zeta)}{2} \ln\left[1 + \frac{D(r_s,\zeta)}{r_s} + \frac{E(r_s,\zeta)}{r_s^{3/2}} + \frac{F(r_s,\zeta)}{r_s^2}\right] + G(r_s,\zeta). \tag{8}$$

Here,

$$D(r_s,\zeta) = e^{-2b_0[\zeta]/a_0[\zeta]} - 2\left(1 - e^{-(r_s/100)^2}\right)\left(\frac{f_0 - c_x(\zeta)}{a_0(\zeta)} + \frac{1}{2}e^{-2b_0[\zeta]/a_0[\zeta]}\right), \tag{9}$$

$$E(r_s,\zeta) = -\frac{2\left(1 - e^{-(r_s/100)^2}\right)f_1}{a_0(\zeta)}, \tag{10}$$

$$F(r_s,\zeta) = e^{-2b_0[\zeta]/a_0[\zeta]} - 2\left(1 - e^{-(r_s/100)^2}\right)\left(\frac{f_2 - c_s(\zeta)}{a_0(\zeta)} + \frac{1}{2}e^{-2b_0[\zeta]/a_0[\zeta]}\right), \tag{11}$$

$$G(r_s,\zeta) = \frac{r_s}{1 + 10e^{(r_s/100)^2} r_s^{5/4}}\left[-a_1(\zeta)\ln\left(1 + \frac{1}{r_s}\right) + b_1(\zeta)\right]. \tag{12}$$

Note that the interpolation ansatz defined in Eqs (8)-(12) are inspired by the work of Chachiyo[13], and we have augmented Eq (3) with more function terms to interpolate the asymptotic limits defined in Eqs (4) and (5). In particular, $D(r_s,\zeta)$ and $E(r_s,\zeta)$ are designed to recover the first two terms in Eq (5), $F(r_s,\zeta)$ is for the interpolation between the $a_0(\zeta)\ln r_s + b_0(\zeta)$ term of Eq (4) and the third term of Eq (5), and $G(r_s,\zeta)$ is employed to recover the $r_s\left[a_1(\zeta)\ln r_s + b_1(\zeta)\right]$ term in Eq (4).

In Eqs (9)-(12), we have used a Gaussian exponential term ($e^{-(r_s/100)^2}$) to control the asymptotic behavior of the functional, following the similar approach employed in the development of the PW91 exchange functional [21]. Moreover, in Eq. (12), there is a prefactor 10 to the Gaussian exponential term in the denominator, which is chosen to regulate the low-density region of $e_c(r,\zeta)$. Note that the factor 100 in the Gaussian exponential term and the prefactor 10 are the intrinsic parts of the chosen function forms, so we do not count them as fitted parameters. Although fitting these two parameters against the QMC data can give slightly improved performance, we chose not to do so.

In the high-density limit $r_s \to 0$, $(1-e^{-(r_s/100)^2}) \to 0$, $D(r_s,\zeta) \to e^{-2b_0[\zeta]/a_0[\zeta]}$, $E(r_s,\zeta) \to 0$, $F(r_s,\zeta) \to e^{-2b_0[\zeta]/a_0[\zeta]}$, and $G(r_s,\zeta) \to r_s\left[a_1(\zeta)\ln r_s + b_1(\zeta)\right]$. Therefore,

$$\begin{aligned} e_c(r,\zeta) &\to -\frac{a_0(\zeta)}{2}\ln\left[1+\frac{e^{-2b_0[\zeta]/a_0[\zeta]}}{r_s}+\frac{e^{-2b_0[\zeta]/a_0[\zeta]}}{r_s^2}\right]+r_s\left[a_1(\zeta)\ln r_s + b_1(\zeta)\right] \\ &\to -\frac{a_0(\zeta)}{2}\ln\left[\frac{e^{-2b_0[\zeta]/a_0[\zeta]}}{r_s^2}\right]+r_s\left[a_1(\zeta)\ln r_s + b_1(\zeta)\right] \\ &= a_0(\zeta)\ln r_s + b_0(\zeta) + r_s\left[a_1(\zeta)\ln r_s + b_1(\zeta)\right] \end{aligned} \qquad (13)$$

Eq. (13) demonstrates that $e_c(r_s, \zeta)$ recovers the high-density limit of Eq. (4).

In the low-density limit $r_s \to \infty$, $(1-e^{-(r_s/100)^2}) \to 1$, $D(r_s, \zeta) \to -\frac{2(f_0 - c_x(\zeta))}{a_0(\zeta)}$, $E(r_s, \zeta) \to -\frac{2f_1}{a_0(\zeta)}$, $F(r_s, \zeta) \to -\frac{2(f_2 - c_s(\zeta))}{a_0(\zeta)}$ and $G(r_s, \zeta) \to 0$,

$$e_c(r_s, \zeta) \to -\frac{a_0(\zeta)}{2} \ln\left[1 - \frac{2(f_0 - c_x(\zeta))}{a_0(\zeta) r_s} - \frac{2f_1}{a_0(\zeta) r_s^{3/2}} - \frac{2(f_2 - c_s(\zeta))}{a_0(\zeta) r_s^2}\right]$$

$$\to -\frac{a_0(\zeta)}{2}\left[-\frac{2(f_0 - c_x(\zeta))}{a_0(\zeta) r_s} - \frac{2f_1}{a_0(\zeta) r_s^{3/2}} - \frac{2(f_2 - c_s(\zeta))}{a_0(\zeta) r_s^2}\right]$$

$$= \frac{f_0 - c_x(\zeta)}{r_s} + \frac{f_1}{r_s^{3/2}} + \frac{f_2 - c_s(\zeta)}{r_s^2} \ . \tag{14}$$

Eq. (14) confirms that the low-density limit, *i.e.* Eq. (5), has been recovered by $e_c(r_s, \zeta)$ of Eq. (8).

Note that the exact form of $b_1(\zeta)$ in Eq. (4) is not known, and only $b_1(0)$ has been determined by Endo *et al.* [16], so the expressions for the spin-unpolarized ($\zeta = 0$) and fully polarized ($\zeta = 1$) cases in W20 are slightly different,

$$e_c^{W20}(r_s, 0) = -\frac{a_0(0)}{2} \ln\left[1 + \frac{D(r_s, 0)}{r_s} + \frac{E(r_s, 0)}{r_s^{3/2}} + \frac{F(r_s, 0)}{r_s^2}\right] + G(r_s, 0), \tag{15}$$

$$e_c^{W20}(r_s, 1) = -\frac{a_0(1)}{2} \ln\left[1 + \frac{D(r_s, 1)}{r_s} + \frac{E(r_s, 1)}{r_s^{3/2}} + \frac{F(r_s, 1)}{r_s^2}\right]$$

$$+ \frac{r_s}{1 + 10e^{(r_s/100)^2} r_s^{5/4}}\left[-a_1(1)\ln\left(1 + \frac{1}{r_s}\right)\right]. \tag{16}$$

As shown in Eq. (16), the $b_1(1)$ term in $G(r_s, 1)$ of Eq. (12) has been removed from $e_c^{W20}(r_s, 1)$, and the final expression for the W20 UEG correlation functional is given as

$$e_c^{W20}(r_s,\zeta) = e_c^{W20}(r_s,0) + [e_c^{W20}(r_s,1) - e_c^{W20}(r_s,0)]f(\zeta), \tag{17}$$

where $f(\zeta)$ is the spin interpolation function developed by von Barth and Hedin [22], and it has been employed in the PZ81 [5] UEG correlation functional,

$$f(\zeta) = \frac{(1+\zeta)^{4/3} + (1-\zeta)^{4/3} - 2}{2^{4/3} - 2}. \tag{18}$$

### III. COMPARISON AND DISCUSSION

Figure 1 shows that the W20 model approaches the high- and low-density limits correctly. In TABLE II, we compare the correlation energy per electron, $e_c(r_s, \zeta)$, calculated from C16 [13], revC16 [18], W20, and the QMC results of Spink et al. [12]. Since Bhattarai et al. [11] have suggested that the QMC results for $r_s = 0.5$ in the work of Spink et al. might not be accurate, only the results for the ranges $1 \leq r_s \leq 20$ are presented in TABLE II. For the $r_s = 1$ ($\zeta = 0, 1$) cases, W20 shows the larger deviation from the QMC results than C16 and revC16, whereas in most of the other cases W20 gives better agreements. In TABLE III, we report the mean signed errors (MSEs) and mean unsigned errors (MUEs) of seven UEG correlation functionals by using the QMC results ($1 \leq r_s \leq 20$) of Spink et al. as references. W20 gives the lowest MUE for $\zeta = 0$, whereas PZ81 is the best performer for $\zeta = 0.34, 0.66$, and 1. If we use average MUE (AMUE) to evaluate the performances of the tested UEG correlation functionals, PZ81 is the best performer, followed by W20, PW92, and VWN80. It is encouraging that the AMUEs of PZ81, W20, and PW92 are less than 0.01 eV. Among the three best-performing functionals, W20 is the only UEG correlation functional developed by the constraint-satisfying approach without any fitting parameters, and is the only UEG correlation functional which recovers the high- and low-density limits. From this perspective, W20 should have broader accuracy than the QMC-data fitted UEG correlation functionals.

As shown in TABLE III, Karasiev's modification (revC16) of C16 reduce the AMUE by 0.007 eV. The performance of DPI is between C16 and revC16, and this might due to the questionable QMC results at $r_s = 1$. If the data for $r_s = 1$ are excluded, the AMUEs for W20, DPI, revC16, and C16 are 0.007 eV, 0.011 eV, 0.013 eV, and 0.020 eV, respectively.

Since Bhattarai *et al.* suggested that PW92 is more accurate than the QMC results for $r_s < 2$, we have calculated AMUEs by using the PW92 data as the references for the range $0.5 \leq r_s \leq 20$, and the AMUEs for W20, DPI, revC16, and C16 are 0.009 eV, 0.010 eV, 0.020 eV, and 0.025 eV, respectively.

As shown in Fig. 2 (a) and Fig. 2(c), at $0 < r_s \leq 2$, W20 and PW92 are very close to each other, and both recover the high-density limits correctly for $\zeta = 0$ and 1. Fig. 2(b) and Fig. 2(d) confirm that PW92 and W20 are close to each other for $2 \leq r_s \leq 40$, but PW92 does not approaching the low-density limits whereas W20 correctly recover the low-density limits for $r_s \geq 50$.

The W20 functional is designed for the correlation energy per electron of the uniform electron gas. In a real DFT calculation, the correlation energy is the integral of $e_c$ ($r_s$, $\zeta$) over hundreds of thousands of grid points in 3-dimensional space including those in the low-density region. Although, on average, W20 agrees with PW92 within 0.01 eV for $0.5 \leq r_s \leq 20$, the integrated correlation energies do have significant differences. As shown in Table IV, the W20 correlation energies for $AlCl_3$, $C_2Cl_6$, and $S_4$ differ from PW92 by 0.6~1 eV, which definitely are not numerical noise.

Since PW92 is popularly employed in the density functionals beyond local spin-density approximation (LSDA), one might think PW92 is more or less the converged form of UEG correlation functional. However, PW92 is semiempirically fitted to the early 1980's Quantum

MC data set of Ceperley and Alder. The accuracy of this QMC data set is not satisfactory and the earlier VWN80 fit does not use all of the Ceperley-Alder data, but only the data for $r_s \geq 10$. Due to the limited precision of the 1980 Ceperley-Alder QMC data, the PW92 fit, even with 9 empirical fitting parameters, does not recover the low-density limit (Eq. (5)). Comparing the UEG correlation energies in Table II of this work to those in Table I of Ref. [11], we can see that the popular QMC-fitted functionals (PW92, PZ81 and VMN80) underestimate correlation energies at $r_s = 10$ and 20 (using the 2013 QMC data of Spink et al. as reference data), and W20 shows much better accuracy. This is because W20 recovers the low-density (high $r_s$) limit correctly whereas the popular QMC-fitted functionals do not. Furthermore, PW92 recovers only the leading $a_0$ and $b_0$ terms in the high-density limit (Eq. (4)), and a recent paper by Ruggeri et al. [23] suggested that PW92 slightly underestimates the UEG correlation energies of the high-density polarized UEG.

Table V summarizes the constraint-satisfaction status and the number of fitted parameters of each tested UEG correlation functional. As shown in Table V, W20 is the only UEG correlation functional which recovers the known high- and low-density limits and does not have any empirical fitting parameters. DPI recovers both limits well but has one parameter derived from QMC data, whereas C16 does not have any fitting parameters but it recovers only the leading $a_0$ and $b_0$ terms in the high-density limit of Eq. (4), and does not recover the low-density limit. As shown in AMUEs of Table III, both DPI and C16 perform worse than W20.

### IV. CONCLUSIONS

Inspired by the recent C16 [13] and DPI [9] models, a UEG correlation functional, W20, is constructed by interpolating the known exact or near-exact high- and low-density asymptotic limits of UEG $e_c(r_s, \zeta)$ using an elegant ansatz without any fitting parameters. The W20 model recovers the high- and low-density limits for ζ = 0 and 1, and the comparative assessments have shown that W20 agrees with QMC and PW92 within 0.01 eV on average for $0.5 \leq r_s \leq 20$. W20 shows better accuracy than the popular QMC-fitted UEG correlation functionals at low-density region, and W20 paves the way for the constructions of the true "first principles" or "*ab inito*" DFT functionals beyond LSDA. Subroutines which calculate the W20 energy density and potential from the spin electron densities are available by email to the corresponding author. We are developing new generalized gradient approximations (GGA) and meta-GGAs based on the W20 UEG correlation functional.

## ACKNOWLEDGMENT

We thank Wuhan University for financial support and computational resources. This work was supported in part by Foshan Xianhu Laboratory of the Advanced Energy Science and Technology Guangdong Laboratory (XHT2020-003) and Fundamental Research Funds for the Central Universities (WUT:2020Ⅲ029, 2020ⅣA100)

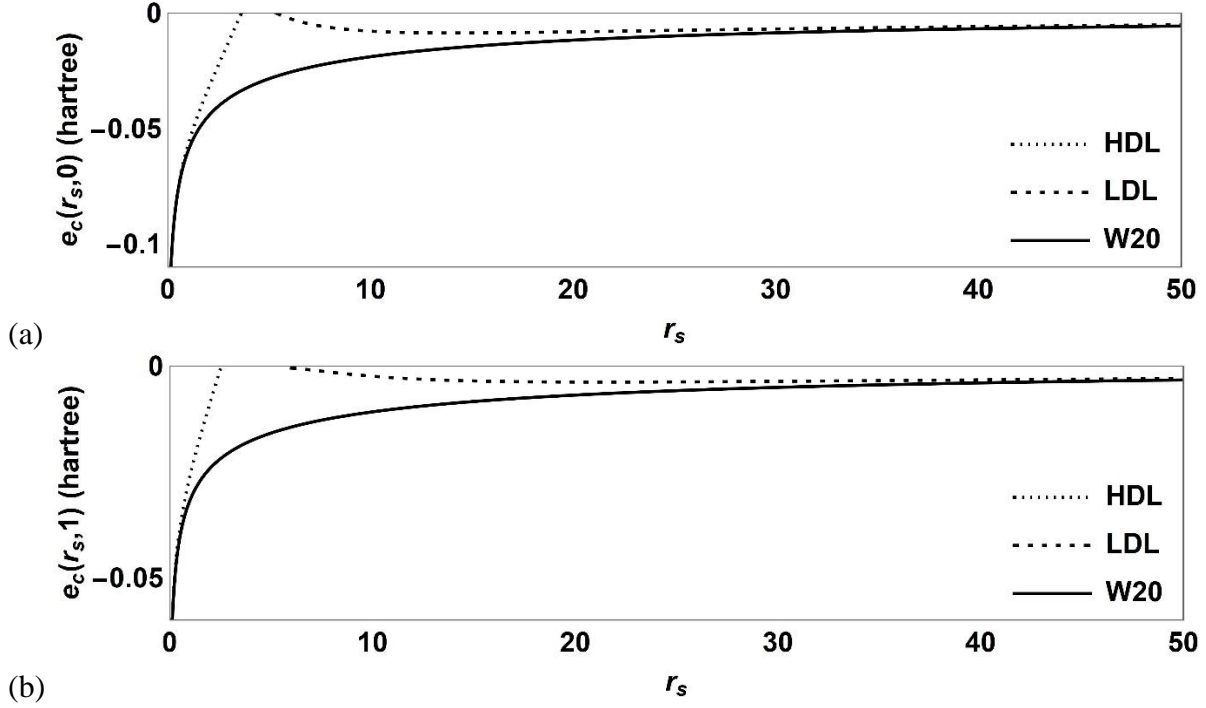

FIG. 1. $e_c(r_s, \zeta=0, 1)$ vs $r_s$ for the high-density limit (HDL, Eq. (4)) and the low-density limit (LDL, Eq. (5)) and the W20 model defined in Eqs. (15), (16), and (17). For $e_c(r_s,1)$, the HDL curve does not have the $b_1(1)$ term.

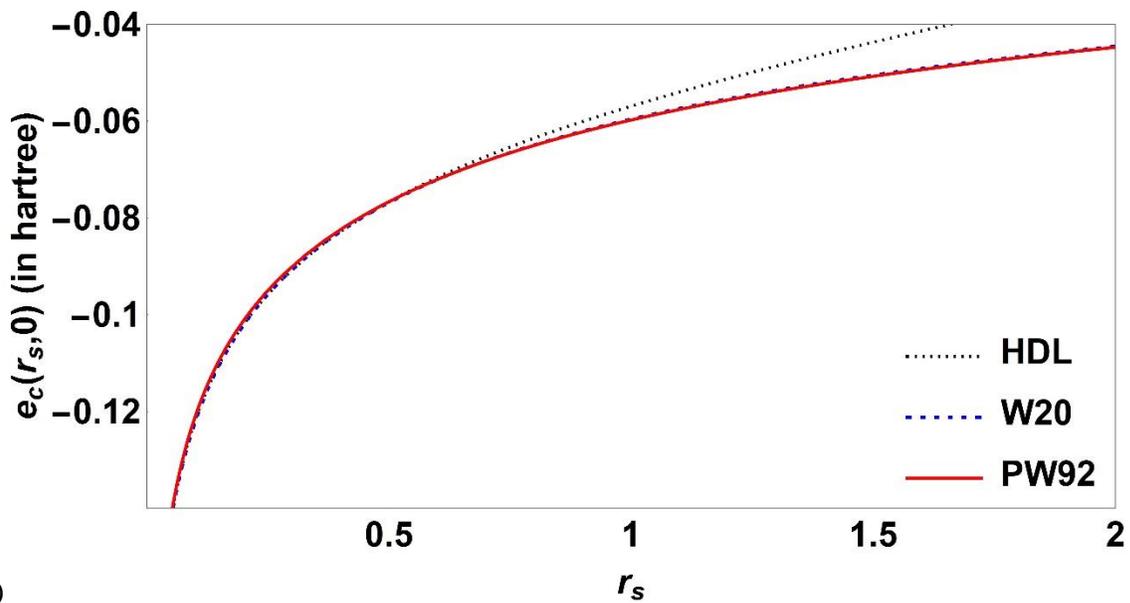

(a)

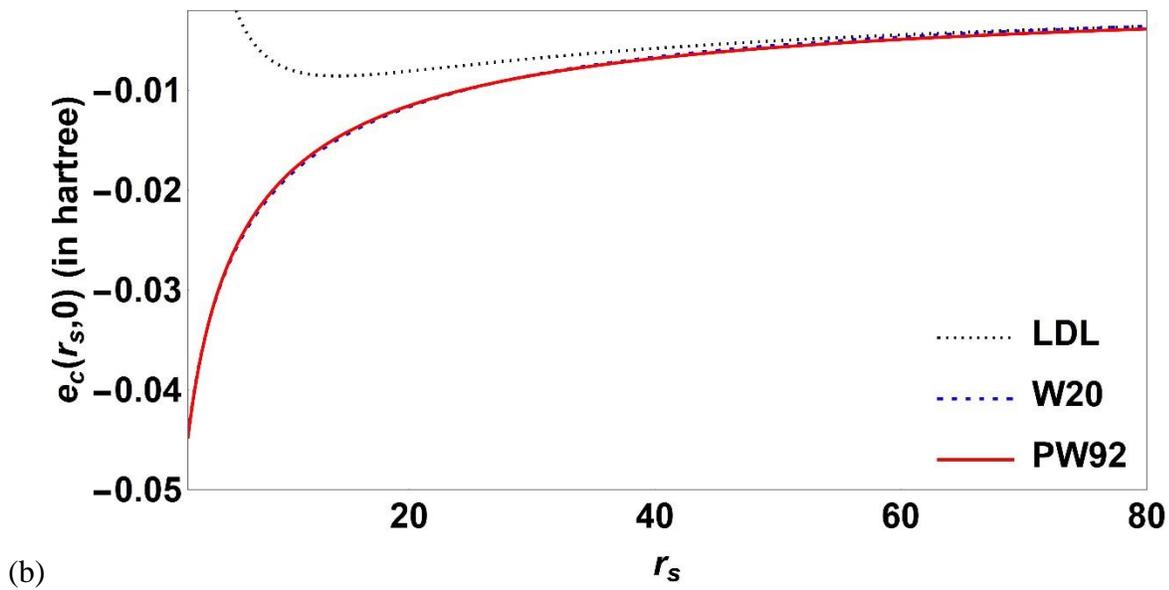

(b)

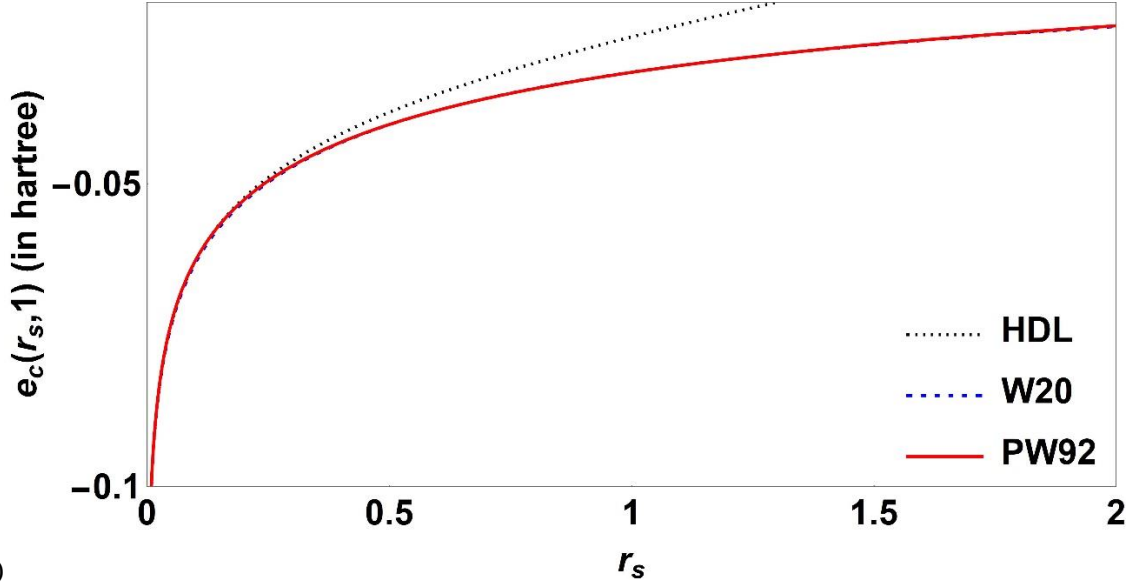

(c)

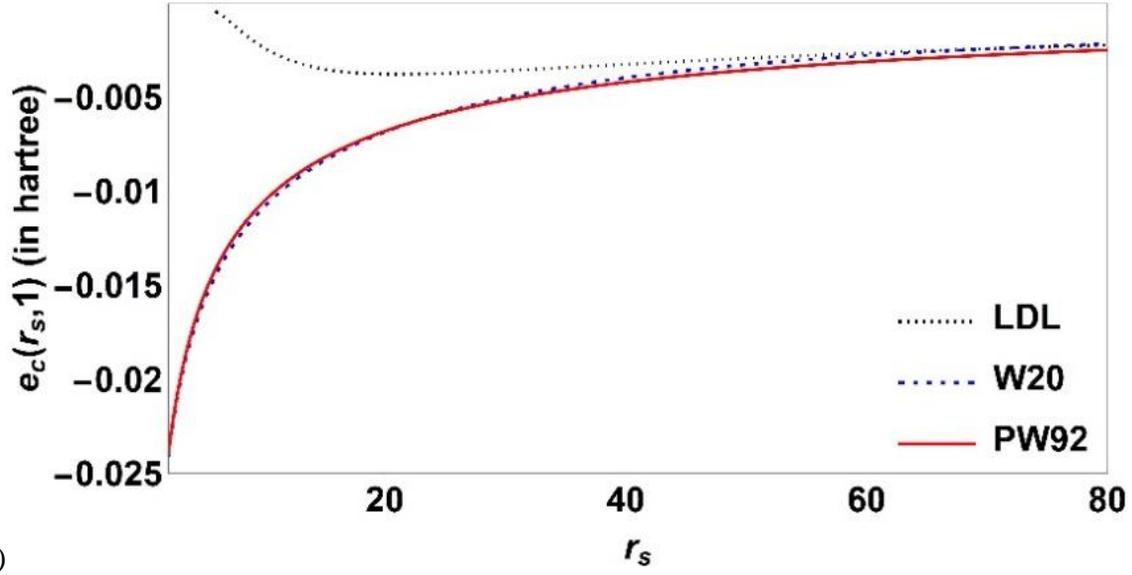

(d)

FIG. 2. Comparison of $e_c(r_s, \zeta)$ vs $r_s$ for W20, PW92, the high-density limit (HDL, Eq. (4)), and the low-density limit (LDL, Eq. (5)): (a) $\zeta = 0$, $0 < r_s \leq 2$; (b) $\zeta = 0$, $2 < r_s \leq 80$; (c) $\zeta = 1$, $0 < r_s \leq 2$; (d) $\zeta = 1$, $2 < r_s \leq 80$.

TABLE I. The exact or near-exact coefficients (in hartree) of the high-density ($r_s \to 0$) and low-density ($r_s \to \infty$) limit (for spin polarization $\zeta=0$ and 1) of the UEG correlation energy in W20. $z(n)$ is the Riemann zeta function. The derivations of the coefficients can be found in the review of Loos and Gill [20]. The coefficients $f_0$, $f_1$ and $f_2$ were taken from Sun, Perdew and Seidl [9].

| | W20 |
|---|---|
| $a_0(0)$ | $(1-\ln 2)/\pi^2$ |
| $b_0(0)$ | $-0.071100 + \dfrac{\ln 2}{6} - \dfrac{3}{4\pi^2} z(3)$ |
| $a_1(0)$ | $(\dfrac{9\pi}{4})^{-1/3} \dfrac{1}{4\pi^3} (\dfrac{7\pi^2}{6} - 12\ln 2 - 1)$ |
| $b_1(0)$ | $-0.01$ |
| $a_0(1)$ | $\dfrac{(1-\ln 2)}{2\pi^2}$ |
| $b_0(1)$ | $-0.049917 + \dfrac{\ln 2}{6} - \dfrac{3}{4\pi^2} z(3)$ |
| $a_1(1)$ | $2^{-4/3} (\dfrac{9\pi}{4})^{-1/3} \dfrac{1}{4\pi^3} (\dfrac{13\pi^2}{12} - 12\ln 2 + \dfrac{1}{2})$ |
| $f_0$ | $-0.9$ |
| $f_1$ | $1.5$ |
| $f_2$ | $0$ |
| $c_x[0]$ | $-\dfrac{3}{4\pi} (\dfrac{9\pi}{4})^{1/3}$ |
| $c_s[0]$ | $\dfrac{3}{10} (\dfrac{9\pi}{4})^{2/3}$ |
| $c_x[1]$ | $-2^{1/3} \dfrac{3}{4\pi} (\dfrac{9\pi}{4})^{1/3}$ |
| $c_s[1]$ | $2^{2/3} \dfrac{3}{10} (\dfrac{9\pi}{4})^{2/3}$ |

TABLE II. Correlation energies per electron (in eV) for UEG from C16 [13], revC16 [18], W20, and QMC data of Spink *et al.* [12].

| $\zeta$ | $r_s$ | Spink *et al.* | C16 | revC16 | W20 |
|---|---|---|---|---|---|
| $\zeta=0$ | 1.0 | -1.605 | -1.580 | -1.593 | -1.621 |
| | 2.0 | -1.218 | -1.182 | -1.198 | -1.212 |
| | 3.0 | -1.010 | -0.978 | -0.995 | -1.003 |
| | 5.0 | -0.774 | -0.752 | -0.769 | -0.772 |
| | 10.0 | -0.510 | -0.499 | -0.515 | -0.513 |
| | 20.0 | -0.316 | -0.309 | -0.321 | -0.317 |
| $\zeta=0.34$ | 1.0 | -1.550 | -1.507 | -1.519 | -1.545 |
| | 2.0 | -1.170 | -1.129 | -1.144 | -1.156 |
| | 3.0 | -0.969 | -0.935 | -0.951 | -0.958 |
| | 5.0 | -0.741 | -0.719 | -0.736 | -0.737 |
| | 10.0 | -0.489 | -0.478 | -0.493 | -0.491 |
| | 20.0 | -0.303 | -0.297 | -0.309 | -0.304 |
| $\zeta=0.66$ | 1.0 | -1.325 | -1.296 | -1.305 | -1.324 |
| | 2.0 | -1.014 | -0.974 | -0.986 | -0.994 |
| | 3.0 | -0.841 | -0.809 | -0.822 | -0.826 |
| | 5.0 | -0.645 | -0.625 | -0.638 | -0.638 |
| | 10.0 | -0.427 | -0.419 | -0.431 | -0.427 |
| | 20.0 | -0.267 | -0.262 | -0.271 | -0.266 |
| $\zeta=1$ | 1.0 | -0.827 | -0.851 | -0.854 | -0.859 |
| | 2.0 | -0.642 | -0.650 | -0.654 | -0.654 |
| | 3.0 | -0.537 | -0.546 | -0.550 | -0.548 |
| | 5.0 | -0.420 | -0.428 | -0.434 | -0.430 |
| | 10.0 | -0.287 | -0.294 | -0.299 | -0.293 |
| | 20.0 | -0.186 | -0.189 | -0.193 | -0.186 |

TABLE III. Statistical summary of errors (in eV) of seven UEG correlation functionals against the QMC results ($1 \leq r_s \leq 20$) of Spink *et al*. [12]. The statistical errors for PZ81, PW92, and DPI were computed by using the raw data in the work of Bhattarai [11]. MSE = mean signed error = mean signed deviation; MUE = mean unsigned error = mean absolute deviation.

| | $\zeta = 0$ | | $\zeta = 0.34$ | | $\zeta = 0.66$ | | $\zeta = 1$ | | AMUE |
|---|---|---|---|---|---|---|---|---|---|
| | MSE | MUE | MSE | MUE | MSE | MUE | MSE | MUE | |
| PZ81 | -0.003 | 0.007 | 0.000 | 0.005 | 0.007 | 0.007 | -0.010 | 0.011 | 0.007 |
| W20 | -0.001 | 0.006 | 0.000 | 0.012 | 0.007 | 0.007 | -0.012 | 0.012 | 0.009 |
| PW92 | 0.000 | 0.007 | -0.007 | 0.013 | -0.006 | 0.010 | -0.007 | 0.008 | 0.010 |
| VWN80 | -0.001 | 0.008 | -0.011 | 0.016 | -0.011 | 0.011 | -0.007 | 0.007 | 0.011 |
| revC16 | 0.007 | 0.010 | 0.014 | 0.015 | 0.011 | 0.014 | -0.014 | 0.014 | 0.013 |
| DPI | 0.005 | 0.015 | -0.008 | 0.025 | -0.002 | 0.018 | -0.005 | 0.012 | 0.018 |
| C16 | 0.023 | 0.023 | 0.028 | 0.028 | 0.022 | 0.022 | -0.010 | 0.010 | 0.020 |

TABLE IV．Correlation Energies (in eV) for three molecules calculated with the W20 and PW92 UEG functionals and the Def2-QZVPP [24] basis set and Hartree-Fock density.

| Molecules | W20 | PW92 |
|---|---|---|
| $AlCl_3$ | -136.38 | -135.79 |
| $C_2Cl_6$ | -240.38 | -239.42 |
| $S_4$ | -135.69 | -135.10 |

TABLE V. Recovering the high-density limit (HDL, Eq. (4)) and low-density limit (LDL, Eq. (5)) and number of fitting parameters of the UEG correlation functionals

| Functionals | HDL | LDL | # of Fitting Parameters |
|---|---|---|---|
| W20 | YES | YES | 0 |
| DPI | YES | YES | 1 |
| C16 | $a_0$, $b_0$ only | NO | 0 |
| revC16 | $a_0$, $b_0$ only | NO | 2 |
| PW92 | $a_0$, $b_0$ only | NO | 9 |
| PZ81 | $a_0$, $b_0$ only | NO | 9 |
| VWN80 | $a_0$, $b_0$ only | NO | 9 |